\documentclass[authorversion,nonacm]{acmart}
\usepackage{tikz}
\usepackage{graphicx}
\usepackage{cleveref}
\usepackage{enumitem}

\newlist{boxlist}{itemize}{2}
\setlist[boxlist]{label=$\square$}
\newlist{circlist}{itemize}{2}
\setlist[circlist]{label=$\bigcirc$}

\def\says#1{``\emph{#1}''}

\setcopyright{rightsretained}

\begin{document}

\title{Computing and Authentication Practices in Global Oil and Gas Fields}

\author{Mary Rose Martinez}
\affiliation{%
  \institution{Brown University}
  \city{Providence}
  \state{RI}
  \country{USA}}

\author{Shriram Krishnamurthi}
\authornote{Partially supported by the US National Science
  Foundation. Author thanks Tiago Guerreiro for citations.}
\orcid{0000-0001-5184-1975}
\affiliation{%
  \institution{Brown University}
  \city{Providence}
  \state{RI}
  \country{USA}}

\begin{abstract}
  Oil and gas fields are a critical part of our infrastructure, and
  vulnerable to attack by powerful adversaries. In addition, these are
  often difficult work environments, with constraints on space,
  clothing, and more. Yet there is little research on the technology
  practices and constraints of workers in these environments. We
  present what we believe is the first survey of oil- and gas-field
  workers located around the world. We establish the presence and
  status of a variety of computing devices and of the security
  practices that govern their use. We also determine the working
  conditions (such as personal protective equipment) under which these
  devices are used, which impacts usable security aspects like
  feasible forms of authentication. We extend these basic insights
  with additional information from a small number of in-depth
  interviews. Our preliminary work suggests many directions for
  improving security in this critical sector.
\end{abstract}

\maketitle

\section{Introduction}

Oil and gas,
a key part of the energy sector and critical infrastructure of many countries, is
a significant security concern~\cite{cisa}. Its core end-products, like petroleum, are
also parts of many other supply-chains: lubricants (``Vaseline'' is petroleum jelly),
plastics, preservatives, artificial limbs,
flame-retardant clothing, and more. Furthermore, oil and gas is
necessarily extracted in low-population areas and transported long
distances, creating many opportunities for attacks.

The oil and gas industry has become increasingly driven by both
information technology (IT) and operational technology (OT). The very
components that make IT susceptible to cyberattack (e.g., software,
data, connectivity, etc.)\ are now found in industrial control systems
(ICS) and process control networks in the OT environment, thereby
vastly increasing a company’s attack surface. Of additional concern
are the potentially grave repercussions of an OT cyber
incident. Unlike an IT cyber incident, an incident in the field or a
manufacturing plant has physical consequences that pose a risk to
human life, safety and/or the environment.

In 2019,
industrial cybersecurity company Dragos~\cite{dragos} said
that the industry
“remains at high risk for a destructive loss of life cyberattack due
to its political and economic impact and highly volatile processes.”
Malware, ransomware, and other attacks on
petrochemical plants and pipelines make the news with some regularity.
According to the 2018 Symantec Internet Security
Threat Report, ICS vulnerabilities increased 29\% from 2016 to 2017~\cite{symantec}.
Furthermore, the Dragos Threat Perspective also assessed that
“state-associated actors will increasingly target oil and gas and
related industries to further political, economic, and national
security goals.” This suggests threat actors who are well-funded and
well-staffed, with more resources at their disposal than the average
cyber criminal.

In this paper, we initiate study of the work environment of oil and
gas personnel. Here, it is important to distinguish the three main
sectors of the industry: upstream, midstream, and downstream.
The \emph{upstream} sector is focused on finding oil and gas
reservoirs, and drilling wells in those reservoirs to extract crude
oil and natural gas. Transportation from reservoirs and storage make
up the \emph{midstream} sector. The \emph{downstream} sector refines
and processes oil and gas into fuels and other materials. (That is,
the product flows from ``up'' to ``down''.)

The contribution of this paper is to
\emph{study oil and gas field
  workers in the upstream sector}. We focus on them for three
reasons. First, they work in locations with great potential for
harm. Second, they often have the most unconventional work
environments of people in this area. Finally, while they share some
similarities to other ``front line'' workers (e.g., medical
personnel), their physical location, bandwidth, etc.\ create unique
challenges. In particular, we try to understand the \emph{computer
  device use} while performing their job:
\begin{itemize}
\item What are the everyday computing practices of upstream oil and
  gas field workers?

\item What ambient factors impact cybersecurity in the oil and
  gas field?

\item Are there any usability challenges?
\end{itemize}

\section{Background on Oil and Gas}

We assume the reader may benefit from a brief primer on the
contemporary oil and gas industry. We
also introduce some terminology useful in the rest of the
paper.

\subsection{The Use of Digital Technology}

While the cyclical nature of the oil and gas industry is the norm,
major shifts have occurred starting from the downturn in 2014~\cite{rogoff}. Its
rapid and protracted nature caused companies to focus on operational
efficiency as oil shifted from over \$100/barrel to under \$65/barrel,
and then under \$40/barrel.  Shifts in both energy generation and
emission reduction are further forcing technology changes~\cite{cghv,d},
a major part of which is the adoption of the Industrial IoT.

\subsection{The Nature of Upstream Workplaces}
\label{s:nature-of-upstream}

\begin{figure*}[t]
\begin{center}
  \includegraphics[height=2.9in]{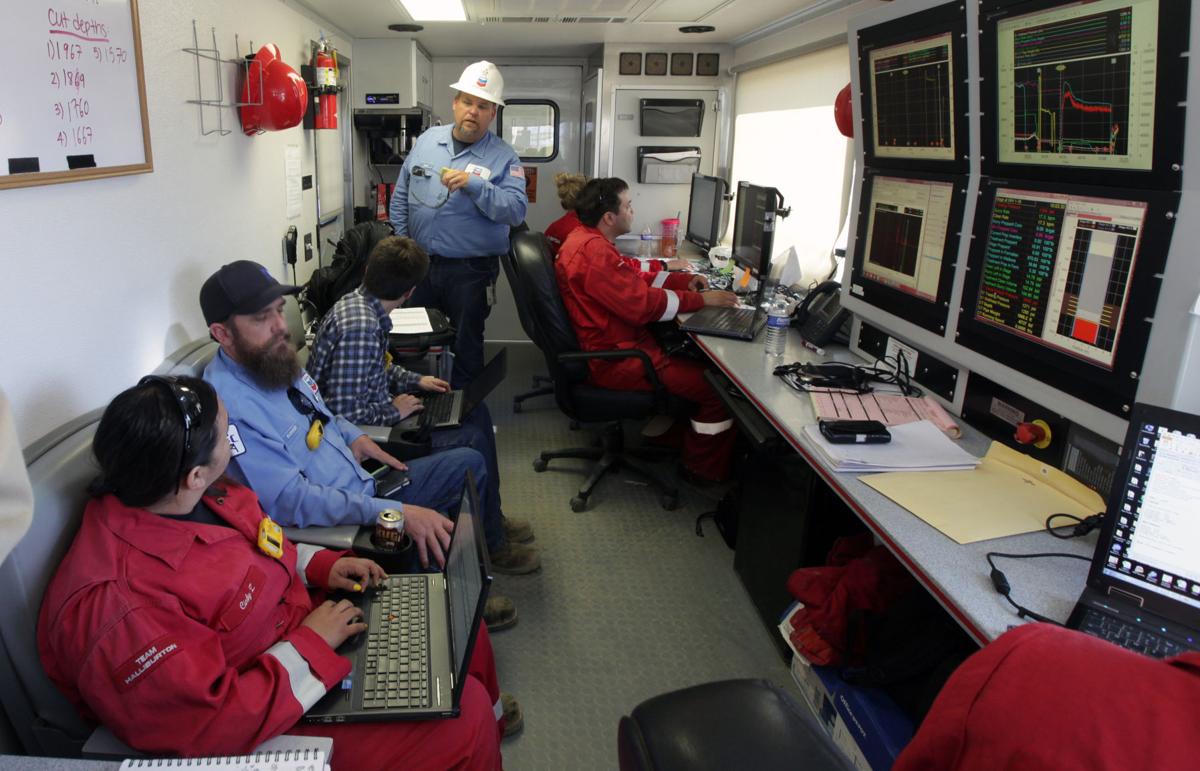}
  \qquad\qquad\qquad
  \includegraphics[height=3in]{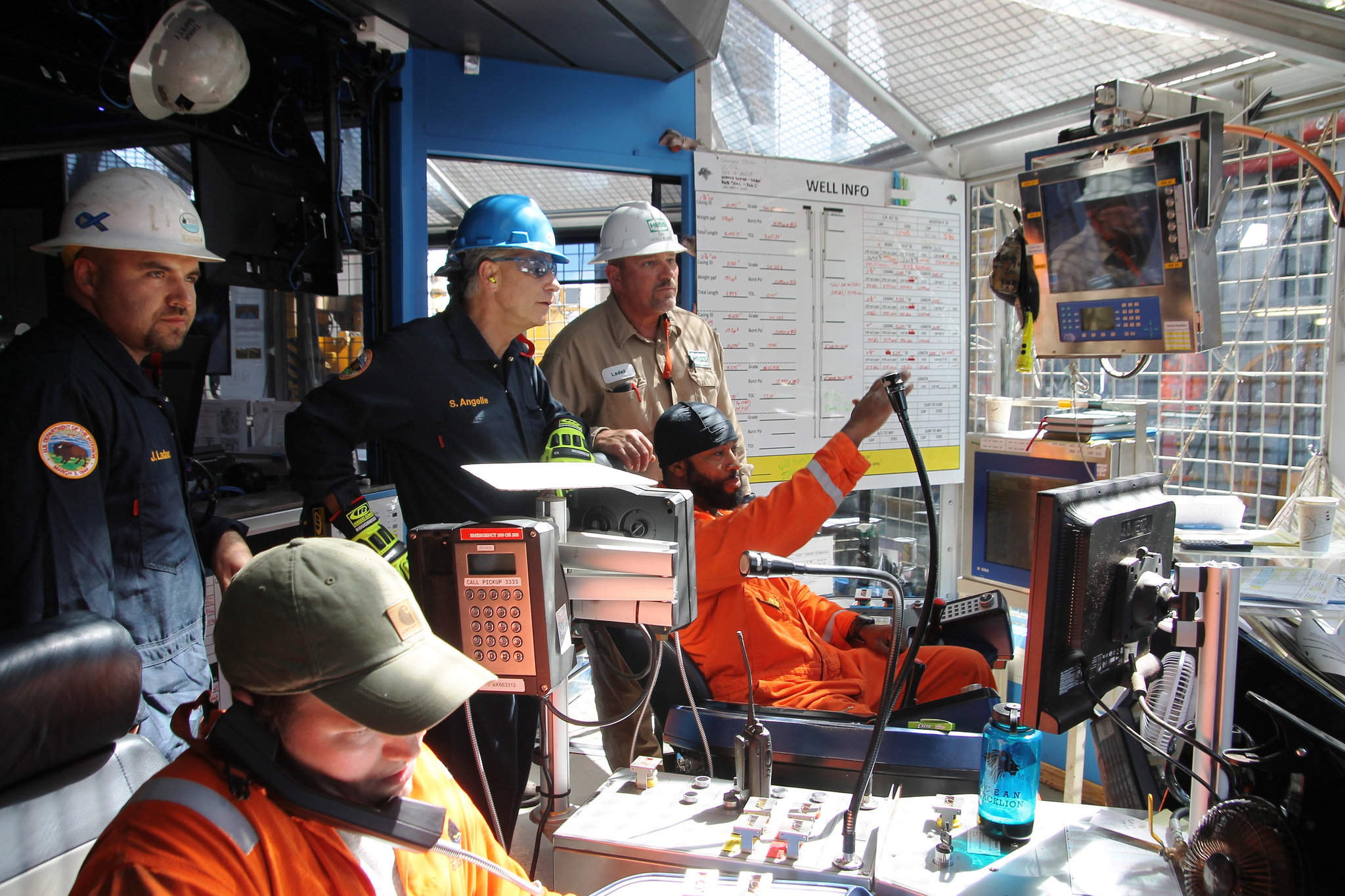}
\end{center}
\caption{Typical onshore (above) and offshore (below) upstream
  workplaces. Images are, respectively, from Felix Adamo / \emph{Bakersfield
    Californian}~\cite{bakersfield-image} (used with permission) and US Bureau of Safety and Environmental
  Enforcement~\cite{bsee} (public domain).}
\label{f:onshore-offshore}
\end{figure*}


Readers may have certain clichéd mental images of the nature of an upstream
oil and gas field. In fact, the upstream workplace may take many
forms, shown visually in \cref{f:onshore-offshore}.

An onshore hydraulic fracturing spread typically includes a data van
where field engineers and other company personnel monitor and manage
the operation. Data vans range in size similar to a recreational
vehicle (RV) and provide an office-like environment in the field. They
have air conditioning, control noise levels, and protect personnel and
computing equipment from the elements. There are desks for 3--4 people
with space for laptops and peripherals such as monitors, keyboards,
and computer mice. Some of the computer peripherals are connected to
rack-mounted personal computers (RMPCs) stored in a small cabinet
along with networking gear to support a local area network on the job
site. For more remote locations without cellular coverage,
connectivity is typically achieved using a satellite dish installed on the
exterior of the van. There may also be a small seating area for guests
who can observe a job using overhead monitors that are connected to
the RMPCs.

Offshore oil and gas fields, on the other hand, are in far-flung and
harsher environments, some located in ultra-deepwater where drilling
occurs in depths greater than 1,500 meters. Their remoteness
means they lack cellular connectivity, and may have much
more drastic connectivity issues (\cref{s:bandwidth}). Compared to
land operations, offshore platforms also have significantly less space
and accommodations for personnel to operate computing equipment.

Unlike in many corporate workplaces, in many upstream sites, there are
many companies with shared governance. Typically, there is an
\emph{operator} who owns the asset (e.g., Shell, Chevron, ExxonMobil);
one or more \emph{service companies} that provide different services
(e.g. Halliburton, Schlumberger, Baker Hughes); and a
\emph{rig owner}. There can actually be several (5--6) service
companies on a particular site.

\subsection{Bandwidth}
\label{s:bandwidth}

In this paper we will see a persistent use of USB devices, which may
surprise some readers, given their known security threats. These have
a strong justification in this industry.

Oil and gas fields generate 
a significant volume of data. Fiber optics monitor a well's
pressure, temperature, and the
flow of fluids as oil and gas.
Nuclear magnetic resonance and pulsed-neutron technology provide
insight into rock formations, such as its lithology and
mineralogy, to find oil and gas reservoirs. Sensors and processors are
used downhole from where tens of thousands of data points every second
are transmitted to drilling engineers who analyze the data to increase
drilling precision in, if possible, near-real time.


In 2017, an offshore platform generated 1-2 terabytes of data daily
while a typical satellite connection provides a transfer rate of 64
Kbps to 2 Mbps~\cite{hand}. The amount of data
generated in an oil or gas field continues to increase with the
utilization of the Industrial IoT. Sensors are used in
a variety of surface operations from monitoring of storage tank levels
to, in combination with analytics and machine learning, preventive and
predictive maintenance of equipment.

Owing to this wide variety of data transmission needs, a
correspondingly wide range of devices is used. Some respondents use USB
drives; others even continue to use USB sticks. Either way, the the
volume of data relative to available bandwidth in effect demands
physical data carriage.

\section{Survey Methodology}

We ran an anonymous, online survey on computing and authentication processes.
The recruitment email is shown in \cref{s:survey-recruit}
and the full survey is in \cref{{s:full-survey}}.
All respondents were volunteers and not compensated.
To protect respondents, an application was submitted
to Brown's Institutional Review Board (IRB). The respondents were
deemed to be Key Informants rather than Human Subjects, and hence not
requiring
IRB approval. Nevertheless, we have applied all standard and
reasonable safeguards in collecting, retaining, and presenting data.

\paragraph{Recruitment}

Using the first author's professional contacts, emails were sent to the
CISOs of several upstream oil
and gas companies to solicit participation from their field
personnel. The email also contained a link to the online survey.

The list of oil and gas companies included some of the largest
operators and oilfield services companies in the world, most of whom
are on the Fortune 500. The invited operators
included both international oil companies and independents, in order
to obtain a global perspective.  Hence, the representation of field
personnel activity and processes spans the various stages and job
types of the upstream oil and gas lifecycle, from drilling to
production.

\paragraph{Procedure}

The online survey was run from early July 2020 to the end of the
month. Invited CISOs were left free
to decide whether to distribute the survey within
their respective companies. We have no evidence that employees were
pressured to respond; had they been, we may have received many
more responses. At one company’s request, the first author
attended global meetings with its OT representatives to provide
context and brief them on the study’s intent. Some companies forwarded
the survey directly. One major company made an internal copy of
it and shared the results; this may have resulted in some redaction,
making our findings an undercount. The survey was conducted anonymously, and
study respondents were not compensated for their time or input.

In addition to demographic information, the survey consisted of three
sets of questions. The first set centered around the computing
equipment used, if any. Secondly, questions were asked regarding
security practices while using computing equipment. Finally, there
were questions geared towards understanding the impact of
ambient factors in field operations. Free-response boxes were
included throughout the survey for optional further elaboration.

Our survey is necessarily limited in scope. First, we were unable to
ask particularly sensitive questions, due to companies wanting to
protect information about vulnerabilities. Second, the survey needed
to be brief: upstream workers often alternate between high-intensity
work on-site followed by periods of relaxation off-site. This makes
their time precious, and did not give us
the luxury of asking about all the security practices of potential
interest. Indeed, we worked extensively to limit our survey size to
persuade the CISOs to share it. Finally, we were unfortunate to run
into COVID-19, which has shrunk the workforce in this sector and has
created significantly greater pressure and job anxiety for those who
remain employed. Our design had to account for these realities.

\paragraph{Respondents}

\begin{figure}[t]

\begin{center}

\begin{tabular}{lrr}
& N & \% \\
\hline
Gender & & \\
\hline
Female & 5 & 6 \\
Male & 79 & 94 \\
\hline
Age & & \\
\hline
18--24 & 2 & 2 \\
25--34 & 24 & 29 \\
35--44 & 23 & 28 \\
45--54 & 24 & 29 \\
55--64 & 10 & 12 \\
\hline
Education & & \\
\hline
High school diploma / secondary education & 5 & 4 \\
Trade / technical / vocational training & 3 & 54 \\
Associate / 2-year degree & 3 & 33 \\
Bachelor's / 4-year degree & 44 & 6 \\
Graduate / 6-year degree or higher & 27 & 4 \\
\end{tabular}
\quad
\begin{tabular}{lrr}
& N & \% \\
\hline
Years of Oil \& Gas Field Experience & & \\
\hline
0-- 5 & 15 & 18 \\
6-- 10 & 19 & 23 \\
11-- 15 & 14 & 17 \\
16--20 & 16 & 19 \\
21--25 & 10 & 12 \\
26--30 & 3 & 4 \\
31--35 & 3 & 4 \\
36--40 & 2 & 2 \\
40+ & 1 & 1 \\
\hline
Work Location (multiple responses allowed) & & \\
\hline
Africa & 3 & 3 \\
Asia & 17 & 20 \\
Central America & 1 & 1 \\
Europe & 18 & 21 \\
Middle East & 20 & 23 \\
North America & 23 & 26 \\
Oceania & 5 & 6 \\
\end{tabular}

\end{center}

\caption{Respondent Demographics.}
\label{f:demographics}

\medskip
\hrule
\end{figure}

A total of 91 people participated in the survey, summarized in
\cref{f:demographics}.
Although the
demographic questions were optional, only 9\% of the respondents
declined to answer. Of the 84 respondents, seventy-nine (94\%) of the
respondents were male and five (6\%) were female. In terms of age
distribution, most were between 25 and 54 years old, with an even
split among the 25–34, 35–44, and 45–54 age ranges.

In terms of education level, in 2017, the RAND Corporation projected
that more than 60\% of jobs from 2014 through 2024 in the upstream
sector will require postsecondary education, a number higher than that
of the midstream and downstream segments~\cite{bgoc}. In comparison, the
majority of respondents had a bachelor (4-year) degree or higher; 33\%
of this group also obtained a graduate degree. Only five people (6\%)
did not have any additional training or education beyond a high school
diploma or secondary education. This indicates that the surveyed group
represents more of the educated upstream workforce than average.

Respondents were also asked about their number of years of experience
in the oil and gas field. The responses reflected a wide range of
experience levels, from 0–5 years of experience to one respondent
having over 40 years of experience. Most people had between 0 and 25
years of experience.

The geographies represented in the survey spanned the globe and
provided relatively good coverage. Multiple responses were allowed in
this question since it is not uncommon for oil and gas field workers
to work in more than one location. The regions most indicated were
North America (26\%), the Middle East (23\%), Europe (21\%), and Asia
(20\%), with one person having worked across all four regions. There
were also five respondents from Oceania and three from Africa. Other
than lower participation from Africa, the geographical distribution is
representative of the oil and gas regions around the world reported in
the 2019 Oil \& Gas Employment Outlook Guide~\cite{ogjs}.

\paragraph{Sampling Threats}

Of course, several factors conspire to skew the respondent
sample. Naturally, this is only a small fraction of the total number
of people employed in this area. Second, they are in a kind of
position where they might be reached by a CISO and respond. Most of
all, the survey was conducted at the height of the COVID-19 pandemic,
at a time when layoffs and job uncertainties were high globally, but
especially so in the oil and gas industries, which had seen demand
worldwide plummet. Some of these factors may explain the higher
education level of our respondents. Nevertheless, despite these
factors, we do not (based on the expert knowledge of an author) find
the responses to the survey especially outside the bounds of
expectation.

\section{Survey Findings}

The overwhelming majority of respondents (96\%) use at least one
computing device at a job site. Only five respondents answered
``none'' to the question “What computing device(s) do you use for work
at a job site?”. (Note that some of the questions allow multiple answers, so
participation may appear to exceed 100\%.)

\subsection{Computing Equipment}

Twenty percent of survey respondents use computing devices that are
fixed (e.g. installed inside a field trailer, mounted on equipment,
etc.). The rest use mobile computers such as laptops and tablets,
some in addition to fixed computing devices. Of the mobile
computing users, 90\% use machines that are assigned to them while
22\% share a computer with others. A mouse and keyboard are the most
commonly used peripheral equipment in the field (85\% use a mouse,
59\% use a keyboard). 15\% also indicated the use of a pen or
stylus.

In terms of external USB devices, 53\% of respondents use USB
storage. Of those, 68\% use one provided by the company and 45\% use a
personal one.

Respondents were evenly split on whether the comfort of the work
environment affected their use of a computer. Some common challenges,
especially for offshore or remote locations, were space restrictions,
limited connectivity, and lack of privacy. On the other hand, a
respondent reported, \says{At the worksite, the computing device is usually
used in a data van, which has the requisite air conditioning and
seating space.} Yet another reported \says{There is a standard office
environment on [a] job site.}

\subsection{Security Practices}

User IDs and passwords remain the prevalent method of logging onto
computers used in the field with 86\% of the computer users. The next
most prevalent was badge scans at 18\%. Biometrics were rare: 6\% use
fingerprints while 4\% use facial recognition. Another 4\% use some
other methods. 18\% used more than one method.

For survey respondents who use a shared computer, a little over
two-thirds use individually assigned user IDs and passwords, while
31\% use a shared user ID and password.

When asked what they \emph{preferred}, support for passwords reduced
to 58\%, with 38\% preferring fingerprints and 23\% facial
recognition. Support for badge-scans was nearly twice its current use,
at 31\% badge scans. 2\% also suggested other methods. Some
readers may find some of these numbers surprising: e.g., why would so
few be in favor of biometrics and so many still in favor of passwords?
We believe the use of protective equipment may have a major impact,
and discuss this in \cref{s:ppe}.

There were a wide range of frequencies with which respondents were
forced to change passwords. The most frequent by far---for 70\%---was,
perhaps surprisingly, every 2--3 months. The next most frequent was
yearly (9\%), every 4--6 months and never (each 6\%), and some (one
person) as frequently as every week. Some also reported more detailed
policies, such as not having to change them unless there was a problem
(\says{all accounts are changed after a
  failed phishing exercise or proof of compromise}). Also, for some
respondents the frequency depended on whether or not multi-factor
authentication was used:
\says{[Passwords are changed] yearly provided you have MFA enabled, 3
  months for non-MFA accounts.}
Respondents also made
distinctions between field computers, corporate computers on the job
site, and control systems.

Given the support that remains for passwords, it is interesting to see
responses to how often they thought passwords \emph{should} be
changed. To researchers, the revised NIST recommendations~\cite{800-63b} to not
demand frequent changes are probably well-known, based on the
rationale that frequent changes lead to weaker passwords~\cite{nist-faq}. Yet the
most chosen response was still every 2--3 months (35\%), followed by
4--6 months (22\%), then monthly (14\%), and annually (12\%),
with 2\% choosing even weekly and daily. Of course, not all
respondents were fans of password changes: one commented that they
should be changed only when passwords are \says{exposed}, another only
\says{when necessary}, and yet another, \says{Changing
  passwords across devices \& applications is time consuming and
  tedious. This should be engineered out.}

One of our interests is in creating better authentication mechanisms
for field workers. Several factors play into this: the ease with which
passwords can be stolen in cramped spaces (which can lead to
misattribution, masking attacks, etc.), the likelihood that
passwords shared between individuals are likely to be of much lower
quality than those kept private, and the interest of respondents in
biometrics. We are therefore curious about the potential to use other
authentication methods (even if not as a primary method, then at least
as a second factor).

\subsection{Personal Protective Equipment}
\label{s:ppe}

While biometrics are attractive,
field personnel do not work in white-collar office space.
Safety is paramount in the oil and gas industry and specific personal
protective equipment (PPE) is required in field operations. Most field
personnel wear coveralls, steel-toed shoes, and hard hats. Additional
PPE may be required depending on the type of operation being
conducted, a person’s job function, and the environment. For instance,
gloves are typically used when managing heavy equipment, ear plugs
protect against high noise levels generated by pumps and other
machinery, and respirators prevent fume inhalation when handling
chemicals. Any and all of these can interfere with one or more forms
of biometric authentication.

This survey was conducted while COVID-19 was well in progress. Our
goal was to get ``steady state'' information that was not overly
biased by temporary factors. We were concerned that simply wording the
survey to indicate that would not be sufficient, since some respondents
might not read the instructions carefully, thereby skewing our
results. We therefore added explicit questions about PPE during
COVID-19 to appear first, followed by the steady-state question,
to make clearer the context when they
reached the general question about PPE.

Both questions we posed to respondents asked ``which personal protective
equipment (PPE) do you use that \textbf{INTERFERES} with your use of
the computer'' (boldface and caps in the original questionnaire).
The survey
had check-boxes for \emph{all} the PPE commonly used in the field,
even if it is a priori unclear how some of them---e.g., steel-toed
shoes---might interfere with computer use.

In both situations, slightly over half indicate that PPE does not
interfere with computer use. Of the remaining respondents, outside
COVID, gloves are the greatest obstruction (64\%), followed by safety
glasses (24\%), masks (19\%), coveralls (14\%), and hard-hats (14\%),
and a few others. Some of the hindrances may not be obvious:
\begin{itemize}

\item \says{A hard hat sometimes `slips', although this may be corrected
  with proper wearing of the hard hat.} This is especially likely to
  occur when leaning over terminals for short-term use.

\item \says{Ear protection needs to be removed when viewing video `guides'
  and `reference materials'.}

\end{itemize}
Under COVID-19, the biggest differences were masks (which went up to
68\% of those reporting interference) and respirators (10\%, as
opposed to nobody outside COVID). Since these are predictable
differences, it suggests that our survey strategy reasonably
distinguishes between COVID-19 and ``steady state'' information.

A small number of respondents may nevertheless have listed all the
PPE they \emph{use} (especially the one who checked all the boxes),
not only that which interferes. We have not, for instance, been able
to determine how steel-toed shoes can be problematic. However, the
reasons may be subtle: e.g., respondents noted that safety goggles can
get smudged or scratched, making it hard to perceive fine-grained details.

Some of these issues, such as slipping hard-hats, seem to be general
computer use issues (and may be better corrected by, e.g., raising the
machine). Others, like ear plugs, may be relevant to the design of
authentication systems (e.g., audio-based authentication). It is
especially important to consider combinations of PPE: for workers
wearing heavy protective gloves, it can be particularly cumbersome to
remove ear plugs.

\section{Interview and Findings}

We followed-up the above survey with a small number of detailed
interviews (full protocol in \cref{s:interview-proto})
 conducted over teleconference. Unfortunately, not all
survey respondents provided us their contact information (which we
solicited, as an optional entry, if they were willing to be
interviewed further). Thus, some respondents whose answers interested
us---e.g., the ones who identified steel-toed shoes as an operational
interference---were not available for clarification.

We did identify four respondents who provided contact information whose
answers we wanted to examine in more depth. They represent two major
companies and three major geographic regions. They were asked about
(a) their computing devices, (b) authentication methods, (c) shared
computer use, (4) USB device use, and (5) PPE use, as well as any
general comments they might have. Below we summarize the salient
portions of these interviews.

\paragraph{P1} uses both individual laptops and shared tablets. They
are primarily used to connect to a rig-control system~\cite{novos}. They are
the only respondent who mentioned the deployment of an access-control
system (in their case, role-based), which gives some people access to
the control systems, others only to maintenance modules, etc. They
are moving away from laptops to tablets (the opposite direction of
P3's preference), because laptops have more moving parts and require
folding. They recommended further hands-free operation through the use
of a voice assistant to give voice commands using a microphone attached to the
throat; they said that noise in the environment can be overcome by
picking up vibrations from the voice box, and had used such a system
earlier in a petrochemical plant.

They use user ID/password authentication. Two-factor is used only
when troubleshooting a rig remotely, and this requires a hard
token. Control systems are also password-protected. They commented
that their primary security concern was that previously un-connected
rigs were becoming connected with increased use of the Industrial
Internet of Things to enable remote control of the wells.

They share tablets, primarily due to budget constraints. There are
also some shared PCs tied to outdated equipment, and are being
retired. A local field office controls passwords, with an internal
audit to confirm that passwords are being altered to company standards.
They personally would have preferred the use of key fob credentials
for unlocking machines, both to get around password changes and
because it would ideally enable them to walk up to use any computer
terminal in the company.

They use various USB devices due to the lack of connectivity. A
critical use is to load a ``well plan'', a set of instructions (e.g.,
maximum RPM) for the control system. USB devices are also used for
maintenance and file transfer (e.g., equipment photos), but these do not
have the same security controls. They use personal USB sticks as
well. All USBs are scanned when plugged into a corporate
laptop. However, only the well plan USBs are subject to extra
scanning. These must hold a ``well key'', a hard-to-guess identifier
that uniquely names a rig/well combination. On insertion, the stick is
checked for a specific file structure with certain files; any
deviation causes the USB to be rejected. Lost USBs are required to be
reported immediately; when this happens, the well key is changed, thus
invalidating the data on the USB.

\paragraph{P2} uses an individual laptop. They currently authenticate
with a user ID and password, with two-factor authentication. They
do not believe that they hold critical information. They find the
password requirements frustrating and prefer biometric authentication;
in particular, they both assume they have no privacy anyway, and do
not believe they hold any critical information. However, they were
unconvinced that badge-based authentication would be a good idea
because badges are lost, forgotten, or destroyed.

They use external USB drives to gather log data (which previously used
to be recorded on DVDs). There is no security on this drive, but the
log data can only be read/interpreted by proprietary software.  They
are also able to use a USB drive to carry personal files, which has
been important as they have relocated to 5 different countries (in
Africa, Asia, and Europe) in 8 years.

\paragraph{P3} uses both an individual laptop and tablets. The tablets
are hardened for field use, but equipped with a stylus, and a groove
to hold it. Their field duties include recording readings/data from
gauges and converting analog to digital data. Though they are equipped
with a hardened tablet, they pointed out that they prefer to use a
laptop because their work requires not only the specialized software
for the above tasks, but also frequent use of Word, Excel, and PDF
viewing. As such, a field-hardened laptop would likely be more useful.

They currently use a user ID/password with two-factor for
authentication. In their opinion, this is preferable to using
biometrics. We found this surprising; their reason is that scanners are
likely to get dirty, and then malfunction. In addition, in their
experience working in an environment where hands easily get greasy,
the scanners do not work even after trying to wipe
hands. (Furthermore, if these hands leave deposits on the scanner,
that would impact future users with clean hands, too.)

Owing to the multiple companies present on site
(\cref{s:nature-of-upstream}), the operator at P3's site actually
allows all the service companies to use the operator's own
laptops. Because many different users may use the same laptop, the
user IDs and passwords are actually written down and posted on the
laptops. (In other evidence that their site may be less digitally
sophisticated that some others, the data capture is sometimes done not
only in Excel but also in Word.)

USBs are used to transmit gauge data to operators, the larger of whom expect
data within 48 hours (smaller operators can expect to wait
longer). These gauges (pressure, temperature, vibration, etc.)\ can
generate over a million data points every hour, but a 12GB USB stick is
sufficient for a job. Thus, there is extensive use of USB
sticks. However, P3 says that the data are useful only to someone who
understands what they are looking at, so USB theft if perceived to be
a low security risk.

Due to connectivity issues, data are also stored on laptops. The
end-to-end work process is now cloud-based, which can be problematic
due to connectivity. When workers are offshore for 1--2 months, data
are recorded on laptops (usually in Excel) and uploaded when back
onshore.

\paragraph{P4} uses an individual laptop. They are not concerned about
theft because ``there is a close community on the rig''. They would
have preferred more apps and connectivity: e.g., they point out that
when there is no connectivity, Hazard Observation Cards/Forms (which
are used in safety engineering) are filled out by hand, which can lead
to lost information either through the loss of the card itself, or
from information lost in the logging process. We note that some of the
incident reports can relate to later security problems: e.g., a
seemingly hazardous behavior could have been a precursor to a later
physical safety attack (especially if there are insider
threats). When this occurs, having a complete and reliable record is
critical.

P4 works with gloves, and does not have the grease problems of
P3. They prefer fingerprint-based authentication, which they find
convenient and perceive to be safe. (They feel that removing gloves is
not a major problem.) They had no privacy concerns with use of
biometric authentication. Interestingly, however, they were dubious
that facial recognition would work, noting that \says{guys look
  different after 30 days on a rig} (a reference to the growth of
facial hair).

P4 does not use a shared laptop, and did not have any problematic
PPE. They also work in an environment that does not use USB
storage. When the upstream site is onshore, they are equipped with a
``data van''; when offshore, they are equipped with satellite
communications (128--256Kbps is common, 512 Kbps at most), which they
anyway need because some jobs require real-time data transfer.

\section{Discussion and Recommendations}

While oil and natural gas are expected to play a role in the global
energy mix for decades~\cite{dnv-gl}, market economics are demanding a marked
change in operating practices to reduce cost. Operators and oilfield
service companies alike are leveraging digital technology in field
operations to increase efficiency and lower operating costs. Digital
technology is further used for real-time monitoring and remote
operations, allowing jobs to be executed by fewer personnel onsite (in
industry parlance, ``de-manning of the rig''). With fewer field hands,
the remaining onsite personnel use technology and software to feed
data to remote operating centers and manage field systems. For
instance, physical checks of fluid levels in a tank are replaced by
liquid level sensors monitored by a computer. This study confirms the
prevalent use of computing in oil and gas field operations with
96\% of respondents confirming that they use
at least one computing device. The results of the study also
underscore the importance of system design for improved usability and
security.

\subsection{Computer Hardware and Software}

Despite the widespread use of mobile devices, 56\% of
respondents still reported using an external keyboard and 80\% still
use a mouse. The widespread use of peripheral equipment indicates that
laptop touchpads and keyboards are not easy to use in the field,
likely due to the required use of gloves---which respondents
consistently reported as interfering with computer use. Weather
conditions may also be a factor (e.g., lack of finger dexterity in
extreme cold; sweaty hands in extreme heat), especially for computers
on fixed equipment that are subject to the elements. One respondent
commented that \says{On the equipment, sometimes it is difficult to use
computers while it is very hot or raining}. The need for additional
peripheral equipment may also exacerbate already existing space
constraints, especially in offshore operations.

Operations managers should consider the increased use of devices with
touchscreens and an attached stylus to alleviate the need for a
mouse. Alternatively, regular work gloves could be replaced with
touchscreen work gloves to further remove friction with device
usage. The user interface of software programs could likewise be
improved to ease the usability of interacting with a
computer. Examples include using button selections instead of dropdown
menus, reducing the need to type information with a keyboard, and
providing support for voice commands and responses. However, the
latter may not be feasible if the device is not located in a tractor
cab, field trailer, or office space where noise levels are more
controlled.

\subsection{Data Transmission}

Half of survey respondents also reported the use of USB storage
devices, a well-known threat vector; comments included the use of USBs
for file transfer and storage. With 45\% of the USBs used in the
survey being personal devices, USBs pose an even more serious risk to
field operations. Even where connectivity is available, it comes with
burdens: \says{Limited internet connectivity offshore means sharing a
  small, cramped room with other individuals to perform duties that
  require connectivity and internet access.}

Given the connectivity constraints in oil and gas field locations,
especially offshore, it may not be possible to eliminate the use of
USB devices altogether. It is not clear to what extent companies have
put in place policies, procedures, and technologies to provide and
allow only company-provided USBs. Our respondents
were not in a good position to describe these policies, and some
companies may not be comfortable sharing this information anyway.

\subsection{Authentication}

The study shows a difference between current and preferred
authentication methods. Survey respondents’ preference for biometrics
or badge scanning would remove the need to use a keyboard and type in
a user ID and password, the current predominant method of
authentication. \says{The quicker the better}, according to one
respondent.

Nevertheless, given the generally perceived cumbersomeness of
passwords, we were surprised that respondents wanted passwords at
all, and wanted them changed frequently (and in some cases, more
frequently than they are now). We failed to ask what password
robustness policies are used, in part to keep the survey tractable and
in part because the respondents may not accurately know. It is
possible that at least some of these systems permit low-quality
passwords. Whether respondents want passwords changed often
\emph{despite} their weakness or \emph{because} of it is unfortunately
a question we cannot answer.

Survey comments indicated the use of individual accounts for corporate
devices but shared accounts for field computers. Hence, a potential
reason for the continued preference of user IDs and passwords could be
that a shared account is required for shared devices and operations
that run longer than a person’s work shift. In addition, it is
possible that individuals are comfortable with using logon credentials
as the customary method, or that there is skepticism around the use
and storage of their biometric information.

The burden accompanying password changes arguably supports the use of
biometrics as an alternate method of authentication for individual
user accounts; badges could be used for shared accounts. Either method
would remove the need to remember and type passwords and allow
stronger security to be implemented at the same time. However, the use
of PPE clearly creates problems in this workplace (gloves being the
PPE most cited as interfering with computer use). Facial recognition
was also chosen by many respondents, but it comes with its own problems
(other forms of PPE like earmuffs and safety glasses; and the growth
of beards). Furthermore, support for touchscreen and/or biometrics may
require upgrades of existing computing equipment which, in turn, has
cost and logistical implications. The same is true for any increased
use of badges, although to a lesser extent.

\section{Related Work}

We are aware of no comparable work on field practices in this sector.
(A partial and abbreviated version of this paper appears in EuroUSEC 2021~\cite{mk:comp-auth-prac-oil-gas-conf}.)
There is other work in security for oil and gas. Some of it
is constructive in nature: e.g., a cryptographic protocol for SCADA
communications that is designed around the low bandwidths
available in critical
infrastructure~\cite{DBLP:conf/acns/WrightKM04}. There are also
investigative studies, including ones cited above, that analyze
attacks. However, we are not aware of ones that focus directly on the
people who actually work in upstream sites.
In terms of authentication and PPE, systems like ZEBRA~\cite{zebra} may be
useful. NIST guidelines for public safety usable security also cover
some PPE-related issues~\cite{NISTIR:8080}. Another rich source of
information is research on situationally-induced impairments and
disabilities, which our domain corresponds to~\cite{siid,10.1145/3319499.3330292}.

\section{Future Work}

There are many directions for future work:
\begin{itemize}
\item Sample a bigger
  population and in more depth.
\item Better understand the security
  mindset of these employees (e.g., do they view USB storage as a
threat?).
\item Investigate secure data transfer for these
  environments.
\item Do on-site work; this is difficult because, even
  if access to a rig were granted, the safety training alone is
  prohibitive (e.g., it includes water survival in case of a helicopter
  crash).
\end{itemize}
Nevertheless, ultimately, a deep understanding of the impact
of PPE and other ambient factors should result in innovation in
both hardware and software design, as well as novel authentication
practices, which seem essential in these constrained domains.

\bibliographystyle{plain}
\bibliography{cites}

\newpage

\appendix

\section{Survey Recruitment}
\label{s:survey-recruit}

Subject: \verb|Brown University Study|

\begin{verbatim}
Due to your current job role, you are invited to take part in a Brown
University research study on how oil and gas personnel use field
computing equipment. The survey will take at most 10 minutes of your
time and the results can help improve field cybersecurity programs.
 
Please complete the following online survey by July 31: [ELIDED]
 
If you have any questions, please contact Mary Rose Martinez at [ELIDED]
\end{verbatim}

\newpage

\section{Full Survey}
\label{s:full-survey}

Options marked $\square$ are multiple-choice while those marked
$\bigcirc$ are mutually-exclusive. All questions were
optional. Several questions had free-response boxes (not shown). One
company internally reproduced the survey; we did not see their version
of it, but their answers suggest it was faithfully reproduced.

\subsubsection*{Computing Equipment}

\begin{itemize}

\item
  What computing device(s) do you use for work at a job site?
  
  \begin{boxlist}
  \item None
  \item Mobile device assigned to me (e.g. laptop, tablet)
  \item Mobile device shared with others (e.g. laptop, tablet)
  \item Fixed computing device (e.g. installed in a field trailer, mounted on
    equipment)
  \item Other (please specify below)
  \end{boxlist}
  
\item
  Think about where you use computing devices while on the
  job site. Does the comfort of the work environment interfere
  with your use of the computer (e.g. weather, air-conditioned
  space)?

  \begin{circlist}
  \item Yes
  \item No
  \end{circlist}

\item
  Do you use a USB storage device/thumb drive as part of your
  job?

  \begin{boxlist}
  \item Company-provided device
  \item Personal device
  \item I do not use a USB storage device.
  \end{boxlist}

\item
  What external equipment do you use with the computer?

  \begin{boxlist}
  \item None
  \item Keyboard
  \item Mouse
  \item Pen/stylus (for tablets)
  \item Other (please specify below)
  \end{boxlist}

\end{itemize}

\subsubsection*{Security Practices}

\begin{itemize}
  
\item
  How do you log onto the computer?

  \begin{boxlist}
  \item User ID and password
  \item Fingerprint
  \item Facial recognition
  \item Badge scan
  \item Other (please specify below)
  \end{boxlist}

\item
  How would you \textbf{PREFER} to log onto the computer?

  \begin{boxlist}
  \item User ID and password
  \item Fingerprint
  \item Facial recognition
  \item Badge scan
  \item Other (please specify below)
  \end{boxlist}

\item
  If you use a shared computer, is there a shared user ID /
  password or does each person have their own user ID /
  password?

  \begin{circlist}
  \item Individual user ID / password
  \item Shared user ID / password
  \end{circlist}

\item
  How often is a password changed?

  \begin{circlist}
  \item With every job
  \item Daily
  \item Weekly
  \item Monthly
  \item Every 2 - 3 months
  \item Every 4 - 6 months
  \item Yearly
  \item Never
  \item Other (please specify below)
  \end{circlist}

\item
  In your opinion, how often should a password be changed?

  \begin{circlist}
  \item With every job
  \item Daily
  \item Weekly
  \item Monthly
  \item Every 2 - 3 months
  \item Every 4 - 6 months
  \item Yearly
  \item Never
  \item Other (please specify below)
  \end{circlist}

\item
  During this period of COVID-19, which personal protective
  equipment (PPE) do you use that \textbf{INTERFERES} with your use
  of the computer?

  \begin{boxlist}
  \item None
  \item Hard hat
  \item Steel-toed shoes
  \item Coveralls
  \item Gloves
  \item Ear plugs or muffs
  \item Safety glasses
  \item Masks
  \item Respirators
  \item Other (please specify below)
  \end{boxlist}

\item
  Under \emph{normal} circumstances (i.e., excluding COVID-19),
  which personal protective equipment (PPE) do you use that
  \textbf{INTERFERES} with your use of the computer?

  \begin{boxlist}
  \item None
  \item Hard hat
  \item Steel-toed shoes
  \item Coveralls
  \item Gloves
  \item Ear plugs or muffs
  \item Safety glasses
  \item Masks
  \item Respirators
  \item Other (please specify below)
  \end{boxlist}

\end{itemize}

\subsubsection*{Demographics}

\begin{itemize}

\item
  What is your gender?

  \begin{circlist}
  \item Male
  \item Female
  \item Other
  \end{circlist}

\item
  What is your age?

  \begin{circlist}
  \item Under 18
  \item 18 - 24
  \item 25 - 34
  \item 35 - 44
  \item 45 - 54
  \item 55 - 64
  \item 65 or older
  \end{circlist}

\item
  What is the highest level of school you have completed?

  \begin{circlist}
  \item Less than high school diploma / secondary education
  \item High school diploma / secondary education
  \item Trade / technical / vocational training
  \item Associates / 2-year degree
  \item Bachelor's / 4-year degree
  \item Graduate / 6-year degree or higher
  \item Other (please specify below)
  \end{circlist}

\item
  How many years of oil and gas field experience do you have?

  \begin{circlist}
  \item 0 - 5
  \item 6 - 10
  \item 11 - 15
  \item 16 - 20
  \item 21 - 25
  \item 26 - 30
  \item 31 - 35
  \item 36 - 40
  \item 40+
  \end{circlist}

\item
  In what region(s) do you currently work?

  \begin{boxlist}
  \item Africa
  \item Asia
  \item Central America
  \item Eastern Europe
  \item European Union
  \item Middle East
  \item North America
  \item Oceania
  \item South America
  \item The Caribbean
  \end{boxlist}

\end{itemize}

\subsubsection*{Contact Information}

If you are willing to be contacted for a brief interview, please
provide your name and the best way to contact you.

[Name]

[Contact Information]

\newpage

\section{Interview Protocol}
\label{s:interview-proto}

\begin{enumerate}
  
\item Regarding the computing device(s) used,
  \begin{enumerate}
  \item What do you use the computing device for?
  \item Do you think there is any cybersecurity risk when using the device?
  \item Is there any additional security needed?
  \end{enumerate}
  
\item Regarding log on methods,
  \begin{enumerate}
  \item Why do you prefer <answer from survey response>?
  \item Do you use multi-factor authentication? If so, which one?
  \item Do you think it is important for access to a computing
    device in the field to be secure?
  \item If you had to choose, would you use the most \emph{convenient}
    method or the most \emph{secure} method? Why?
  \end{enumerate}

\item If using a \emph{shared} computer and credentials,
  \begin{enumerate}
    \item Why is sharing needed (e.g., operational disruption, shift
      work, easier, etc.)?
    \item Who sets/changes the password? How is it shared with
      everyone who needs it?
    \item Do you have any security concerns regarding this? If so, is
      there anything that can be done to mitigate some of the security
      risks?
  \end{enumerate}

\item If using a USB drive,
  \begin{enumerate}
    \item What is the device used for?
    \item Do you think there are any cybersecurity risks with using USBs?
    \item If the USB drive is used to store and/or transfer data,
      \begin{enumerate}
        \item Who can access and use the data that has been stored?
        \item Do you think the data is adequately protected? If not, what else can be done?
        \item Is the USB drive prepared for reuse for another job
          (e.g., data is wiped)?
      \end{enumerate}
    \item Do you use a USB for all jobs or only certain jobs (e.g.,
      depending on job type, field location, available connectivity,
      etc.)?
    \item What changes can be made so USBs are no longer needed?
    \item Does your computer allow any USB or only company-authorized
      USBs to be connected?
    \item For computers that accept any USB, do you use a personal USB
      or one provided by the company?
      \begin{enumerate}
      \item Do you think one is safer than the other?
      \item If using a personal USB, would you use a company USB if
        the company provided one?
      \end{enumerate}
  \end{enumerate}

\item If any PPE interferes with the use of the computer,
  \begin{enumerate}
    \item How does it interfere?
    \item What would make it easier for you to use the computer while
      keeping in mind the need for PPE?
  \end{enumerate}

\item Are there any other challenges with using computing devices in the field that we have not discussed?

\item Are there any other cybersecurity concerns in the field that we have not discussed?

\end{enumerate}

\end{document}